\documentclass[prl,twocolumn,superscriptaddress,preprintnumbers,showpacs,byrevtex,floatfix,stfloats,float,wrapfig]{revtex4}
\usepackage[pdftex]{graphicx}
\usepackage{amssymb}
\usepackage{amsmath}
\usepackage{multirow}
\usepackage{graphicx}
\usepackage{hyperref}
\usepackage[english]{babel}

\begin{document}
\DeclareGraphicsExtensions{.pdf,.png,.jpg,.eps,.tiff}
\title{Doppler-free Yb Spectroscopy with Fluorescence Spot Technique}
\author{Altaf H. Nizamani}

\author{James J. McLoughlin}

\author{Winfried K. Hensinger\footnote{Electronic address:  W.K.Hensinger@sussex.ac.uk\\
URL:  http://www.sussex.ac.uk/physics/iqt}}

\affiliation{Department of Physics and Astronomy, University of Sussex, Falmer, Brighton, East-Sussex, BN1 9QH, United Kingdom\\
}

\begin{abstract}
We demonstrate a simple technique to measure the resonant frequency of the 398.9 nm $^{1}S_{0}\leftrightarrow{^1}P_{1}$ transition for the different Yb isotopes. The technique, that works by observing and aligning fluorescence spots, has enabled us to measure transition frequencies and isotope shifts with an accuracy of 60 MHz. We provide wavelength measurements for the transition that differ from previously published work. Our technique also allows for the determination of Doppler shifted transition frequencies for photoionisation experiments when the atomic beam and laser beam are not perpendicular and furthermore allows us to determine the average velocity of the atoms along the direction of atomic beam.
\end{abstract}
\maketitle
\setcounter{secnumdepth}{1}
\section{Introduction}
Ytterbium, both atomic and singly ionised, is an element widely used in experiments involving trapped atoms and ions, such as laser cooling and trapping of neutral atoms and ions \cite{Klein,Lehmitz,Roberts,Enders,Honda2,Rapol,Huesmann,Edwards,DAS2,Loftus2,Maruyama,Tamm}, atomic clocks \cite{Barber,Porsev,Hong,Barber2}, frequency standards \cite{Park,Roberts2,Blythe,Kjargaard}, quantum computing experiments \cite{Monroe,Balzer}, quantum optics \cite{Bacon} and atomic parity non-conservation experiments \cite{DeMille}. Knowledge of the $^{1}S_{0}\leftrightarrow^{1}P_{1}$ transition line in atomic Yb and corresponding frequency shifts for the stable isotopes is very important in these experiments as they allow for laser cooling and isotope selective photoionisation \cite{Kjargaard,Hendricks,Tanaka,Balzer,braun}. Various methods have been used to investigate these transitions, and corresponding isotope shifts \cite{meggers,Das,Loftus,Deil,Zinkstok}.\\
We present our measurements of the ytterbium $^{1}S_{0}\leftrightarrow{^1}P_{1}$ transition frequencies which differ from previously published results \cite{Das}. Using a simple method based upon observing and aligning fluorescence spots of atomic Yb generated from an atomic oven in an evacuated glass beljar we have measured resonant transition frequencies and isotope shifts to an accuracy of 60 MHz. Our results are compared to the results obtained using saturation absorption spectroscopy and were further verified by ionising and trapping Yb ions.\\
The advantage of our technique is that it can be used to determine resonance frequencies when the atomic beam is not necessarily perpendicular to the propagation of the laser, which can be the case in many experiments. These Doppler shifts also depend on the mean velocity of the atomic beam, so the measurement is quite useful. Using our technique we were able to measure the Doppler shift of the transition frequency under a range of angles and then derive average velocity of atoms along the direction of atomic beam.
\section{Experimental Setup}
To drive the $^{1}S_{0}\leftrightarrow{^1}P_{1}$ 399 nm dipole transition an inexpensive in-house built ultra violet external cavity diode laser (ECDL) was used. The ECDL was formed with a
laser diode (frequency selected to 399 nm, Sanyo diode: DL-4146-301S) and a diffraction grating (Newport: 53-*-240R) positioned in the Littrow configuration. A piezo electric actuator was
used to adjust the angle of the grating and thus tuning of the emission wavelength. To stabilise the emission wavelength a High Finesse wavelength meter (WS-07, with specified wavelength accuracy of 60 MHz) and NI-DAQ cards operating under labVIEW real time were used. The wavemeter is calibrated using a 780 nm laser which is  locked to within 1 MHz of the $^{87}$Rb D$_{2}$ line. It measures the wavelength of the ECDL and sends the information to a labVIEW program, via a NI-DAQ Card (NA-6143). The program determines a drift in wavelength and generates the corresponding error signal to keep the laser stable.\\
To perform saturation spectroscopy on the $^{1}S_{0}\leftrightarrow{^1}P_{1}$ 399 nm isotope transitions the ECDL frequency was required to scan over a range of 3 GHz. However, adjustment of
the grating angle alone in the ECDL resulted in a mode hop free tuning range of only 800 MHz. To overcome this the piezo voltage and laser diode current were adjusted simultaneously to
produce a mode hop free tuning range of over 5.5 GHz.
For our technique an atomic flux propagating in a known direction was required as this enabled localised fluorescence spots that could be be spatially resolved. To generate the Yb atomic
beam a resistivly heated stainless steel oven tube (2 cm long and 1.5 mm diameter) was used. A small piece of natural Yb was placed inside the oven and Yb atomic beam could be obtained by
running an electric current of 5 to 6 amperes over the oven. The atomic oven was placed inside a vacuum chamber maintained at a pressure on the order of $10^{-8}$ torr.

\section{Fluorescence spot technique}

\begin{figure}[!t]
\centering
\includegraphics[scale=0.16] {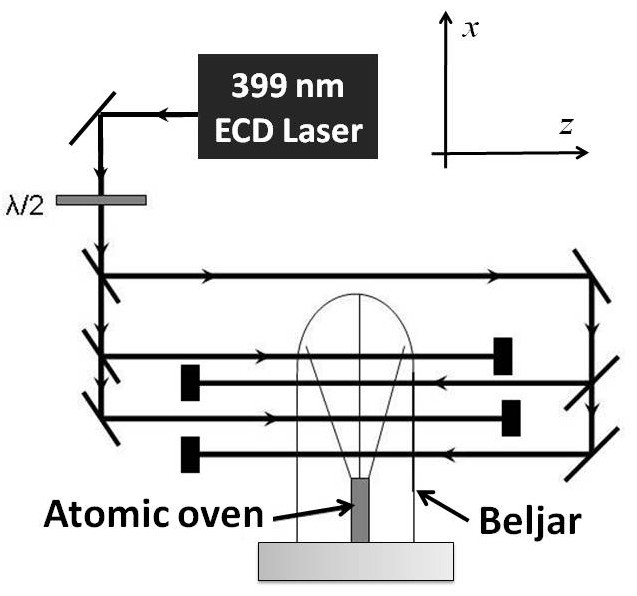}(a)
\includegraphics[scale=0.15] {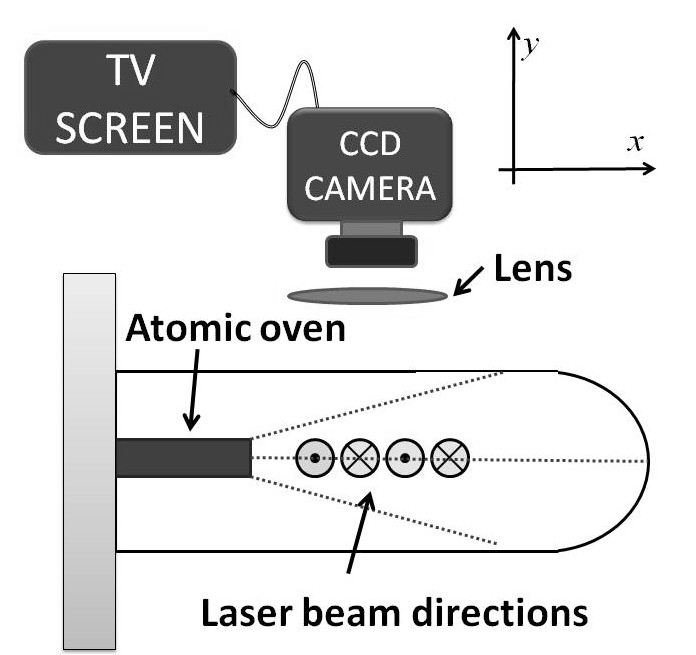}(b)
\caption{Schematic diagram (a) shows the counter propagating laser beams crossing the atomic beam and (b) shows the direction of laser beams and TV-camera setup.}
\label{fig:scheme}
\end{figure}

\begin{figure}[!h]
  \centering
 \includegraphics[scale=0.128]{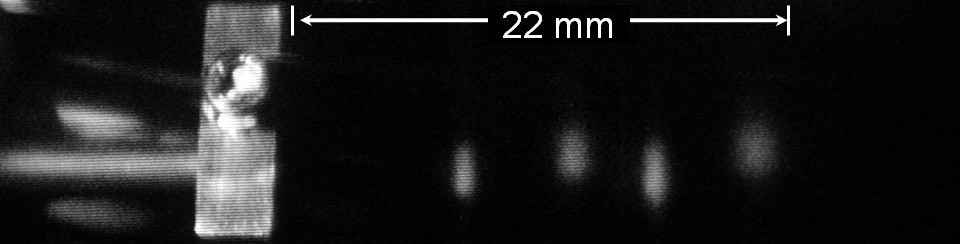} (a)
 \includegraphics[scale=0.105]{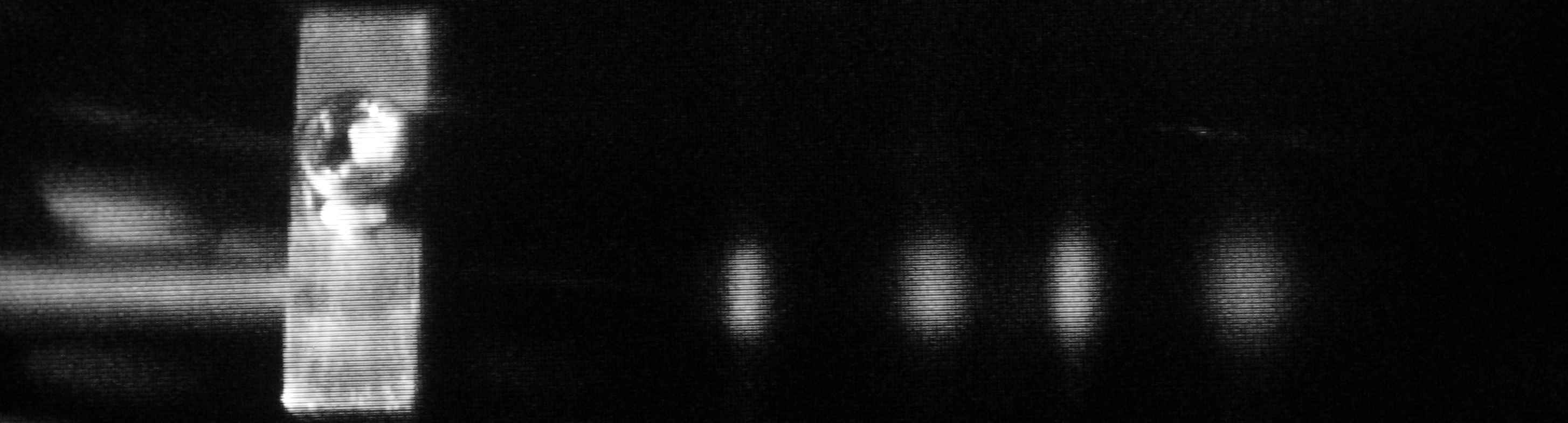} (b)
\includegraphics[scale=0.1]{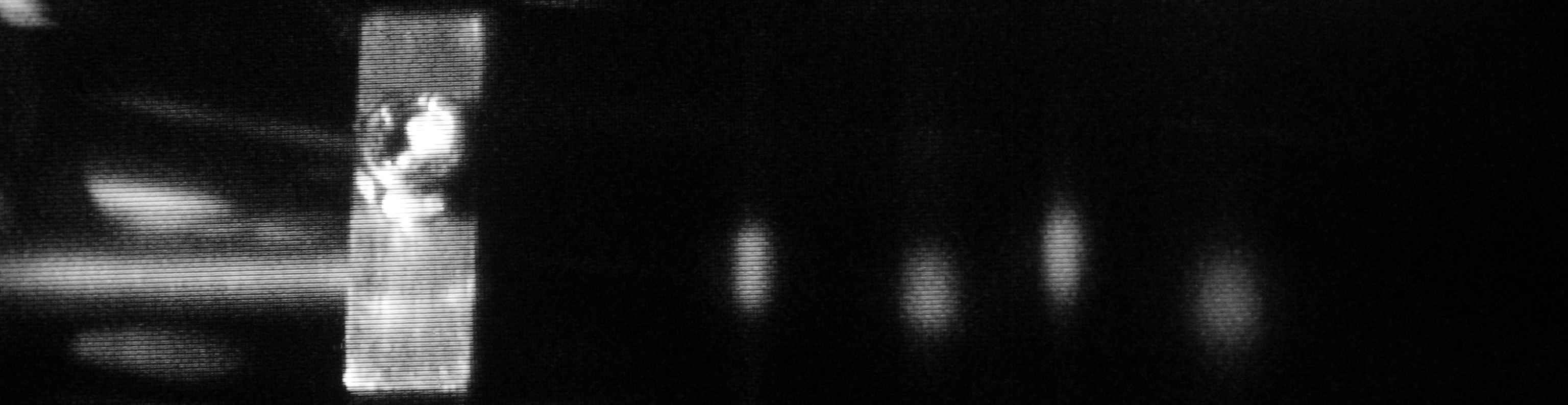} (c)
\caption{Illustration of the resolution of the spectroscopy method. Pictures show the atomic ovens and fluorescence spots where the laser beams intersect the atomic beam taken, (b) on
resonance, (a)-20MHz and (c)+20MHz detuning.}
\label{fig:illustration}
\end{figure}

 We have devised a simple method that works by observing fluorescence spots produced by laser-atom interaction at resonant wavelengths. The experimental setup for our spot method can be
 seen in figure \ref{fig:scheme}, here two pairs of non-overlapping counter propagating laser beams (each with an intensity of approximately 4.5  mWcm$^{-2}$) are passed through an atomic beam. For orientation the cylindrical axis of the oven tube is considered as the reference axis and the lasers are aligned perpendicular to this reference axis unless otherwise stated.\\
The atoms from the oven interact with the four laser beams producing 4 fluorescence spots. In the frame of reference of laser beam propagation each laser interacts with atoms of the same
velocity group. When viewed in the laboratory frame the four spots are seen to be misaligned when the frequency is detuned from resonance, but when on resonance the four spots are seen to
align perpendicular to the propagation of the laser beams. In this situation each laser is interacting with atoms of zero velocity perpendicular to the atomic beam. Spot alignment can be seen in figure \ref{fig:illustration} where \ref{fig:illustration}(b) corresponds to when the laser frequency is on resonance and \ref{fig:illustration}(a)/(c) corresponds to when the laser frequency is detuned -20 MHz/+20 MHz from resonance respectively. This phenomenon occurs irrespective of the angle between the cylindrical axis of the ovens and the laser beams, which is discussed later.\\
In order to obtain the $^{1}S_{0}\leftrightarrow {^1}P_{1}$ transition frequencies, the frequency of the 399 nm laser diode was scanned slowly over 3 GHz. During the scan the florescence spots were imaged using an inexpensive CCD camera setup. After narrowing down the scanning range, the frequencies of the transitions of the different isotopes were
individually measured by aligning the four spots in a straight line with the cylindrical axis of the atomic oven. This is shown in figure \ref{fig:illustration}. We have measured the frequency of the $^{174}$Yb$^{1}S_{0}\leftrightarrow {^1}P_{1}$ transition to be 751.52665 THz $\pm$60 MHz which disagrees with a previously published value 751.525987761 THz $\pm$60 kHz \cite{Das} by approximately 670 MHz. Das \emph{et al.} \cite{Das} relied in their measurement on the stated accuracy of 20 MHz of a home built wavemeter and we speculate that the uncertainty of that wavemeter had been larger than anticipated by the authors. Our measurement differs from the NIST Atomic Spectra Database \cite{NIST} by 260 MHz, which lists the energy of the Yb $^1$P$_1$ level at 25068.222 cm$^{-1}$ (751.52639 THz). Closer examining this discrepancy, the database lists Meggers \emph{et al.} \cite{meggers} as the most recent source of their data. However, Meggers \emph{et al.} \cite{meggers} lists the level energy at 25068.227 cm$^{-1}$ (751.52654 THz) which is only 110 MHz away from our measurement, pointing to a possible typographical error on the NIST Atomic Spectra Database. Considering that Meggers \emph{et al.} \cite{meggers} used a natural mixture of Yb isotopes, their result is consistent with our measurement. We also note that the result by Meggers \emph{et al.} \cite{meggers} shows an expected 550 MHz discrepancy with the measurement by Das \emph{et al.} \cite{Das}.
Using our measurements the isotope shifts relative to the $^{174}$Yb $^{1}S_{0}\leftrightarrow{^1}P_{1}$ transition were then calculated. Our results are shown in Table: \ref{tab:result} along with recently published results. The frequency shift between the $^{172}$Yb $^{1}S_{0}\leftrightarrow{^1}P_{1}$, $^{173}$Yb $^{1}S_{0}(F=1/2)\leftrightarrow{^1}P_{1}(F=3/2)$, and $^{173}$Yb $^{1}S_{0}(F=1/2)\leftrightarrow{^1}P_{1}(F=7/2)$ transitions is 55 MHz, while the separation between the $^{170}$Yb $^{1}S_{0}\leftrightarrow{^1}P_{1}$ and the $^{171}$Yb $^{1}S_{0}(F=1/2)\leftrightarrow{^1}P_{1}(F=1/2)$ transitions is 38 MHz \cite{Das}. These close transitions result in overlapping of the fluorescence spots and cannot be resolved by spot method. The isotope shifts data presented in Table: \ref{tab:result} for the Yb $^{1}S_{0}\leftrightarrow{^1}P_{1}$ transitions are in very good agreement with previously published work \cite{Das,Loftus,Deil,braun}.

\begin{table*}[!t]
\caption{The frequency shifts for the various isotopes of Yb from the $^{174}$Yb $^{1}S_{0} \leftrightarrow{^1}P_{1}$ transition line. The values are obtained from the fluorescence spot technique and compared with the previous published work.}
\begin{tabular}{ l*{5}{c}c }
   \hline
    \hline
  Shift from $^{174}$Yb (MHz)\\ \hline
   Isotope     & Transition            & This work    &    Ref.\cite{Das}        & Ref.\cite{Loftus}    &      Ref.\cite{braun}       &Ref.\cite{Deil} \\

   \hline
\hline
   168       & $^{1}S_{0}\leftrightarrow{^1}P_{1}$     & 1883 $\pm$30  & 1887.400 $\pm$0.05&                &             &1870.2 $\pm$5.2 \\\hline
   170$^{*}$ &  $^{1}S_{0}\leftrightarrow{^1}P_{1}$    &1149 $\pm$60  & 1192.393 $\pm$0.066& 1175.7 $\pm$8.1 & &1172.5 $\pm$5.7 \\
   171$^{*}$ & $^{1}S_{0}(F=1/2)\leftrightarrow{^1}P_{1}(F=1/2)$ &   "         &1153.696 $\pm$0.061&1151.4 $\pm$5.6&1104$\pm$69&1136.2 $\pm$5.8\\\hline
   171       & $^{1}S_{0}(F=1/2)\leftrightarrow{^1}P_{1}(F=3/2)$ & 829 $\pm$30  & 832.436 $\pm$0.05  & 832.5 $\pm$5.6 & 822 $\pm$51  & 834.4 $\pm$4.0  \\\hline
   172$^{*}$ &$^{1}S_{0}\leftrightarrow{^1}P_{1}$      & 546 $\pm$60   & 533.309 $\pm$0.053 & 527.8 $\pm$2.8 & 534$\pm$33 &                \\
   173$^{*}$ & $^{1}S_{0}(F=1/2)\leftrightarrow{^1}P_{1}(F=3/2)$ &"              & 515.972 $\pm$0.2   &                &    \\
   173$^{*}$ &  $^{1}S_{0}(F=1/2)\leftrightarrow{^1}P_{1}(F=7/2)$&"              &587.986 $\pm$0.056  &578.1 $\pm$5.8&\\\hline
   173       &   $^{1}S_{0}(F=1/2)\leftrightarrow{^1}P_{1}(F=5/2)$& -264 $\pm$30 & -253.418 $\pm$0.05 &                  & -262 $\pm$16    \\\hline
   176       & $^{1}S_{0}\leftrightarrow{^1}P_{1}$      & -509 $\pm$30 & -509.310 $\pm$0.05 & -507.2 $\pm$2.5  &  -554 $\pm$35  &                 \\
   \hline
    \hline

\end{tabular}\\
\emph{* can not be resolved with the spot method}
\label{tab:result}

\end{table*}

\subsection{Error analysis}
To determine the resonant frequencies and isotope shifts we used the WS7 wavemeter from High Finesse (specified to 60 MHz accuracy). The wavemeter is calibrated using a 780 nm laser which is  locked to within 1 MHz of the $^{87}$Rb D$_{2}$ line. A He-Ne laser (SIOS SL02/1 calibrated to 1 MHz) was then used to provide confirmation of the calibration. A further two point check was performed to ensure calibration in the ultra violet frequency range. Using our frequency doubling system, that converts 739 nm light into 369 nm, we measured the two wavelengths simultaneously and confirmed the wavemeter to operate within specifications. The wavemeter was calibrated before and after measurements and no change in calibration measurements were observed. The wavemeter is specified to have an accuracy of 60 MHz for absolute frequency measurements, but is more accurate for relative frequency measurements of closely spaced transitions. For the 399 nm transitions, the relative accuracy is translated to be 20 MHz. Therefore the error on transition frequencies is 60 MHz and the error on isotope shifts is 30 MHz. However due to the actual line width of 20 MHz of these transitions and resulting overlap of some transition lines, isotope shift could not be resolved better than 60 MHz.\\
Other sources of error include non-parallel alignment of the lasers where beam misalignment of 1 degree related to 15 MHz error on wavelengths. As the technique can resolve a 20 MHz detuning from resonance it was possible to align the lasers to better than 1 degree as misalignment would result in a visible deviation of the spots. Power broadening of the transition line was insignificant as each beam was less than $4.5\pm 0.9$ mWcm$^{-2}$ whereas the saturation intensity of the transition is 60 mWcm$^{-2}$, which translates to a 3 MHz linewidth broadening.
The potential resolution of the technique can be seen from figure \ref{fig:illustration} where the frequency of the laser has been detuned from (b) resonance by (a)-20 MHz and (c)+20 MHz. The separation from the front of the oven to the furthest spot is approximately 22 mm, and by observing spot misalignment at a given detuning we extrapolate that the resolution of the technique can be 10 MHz in this particular setup.

\begin{figure}[!b]
\includegraphics [width=80mm] {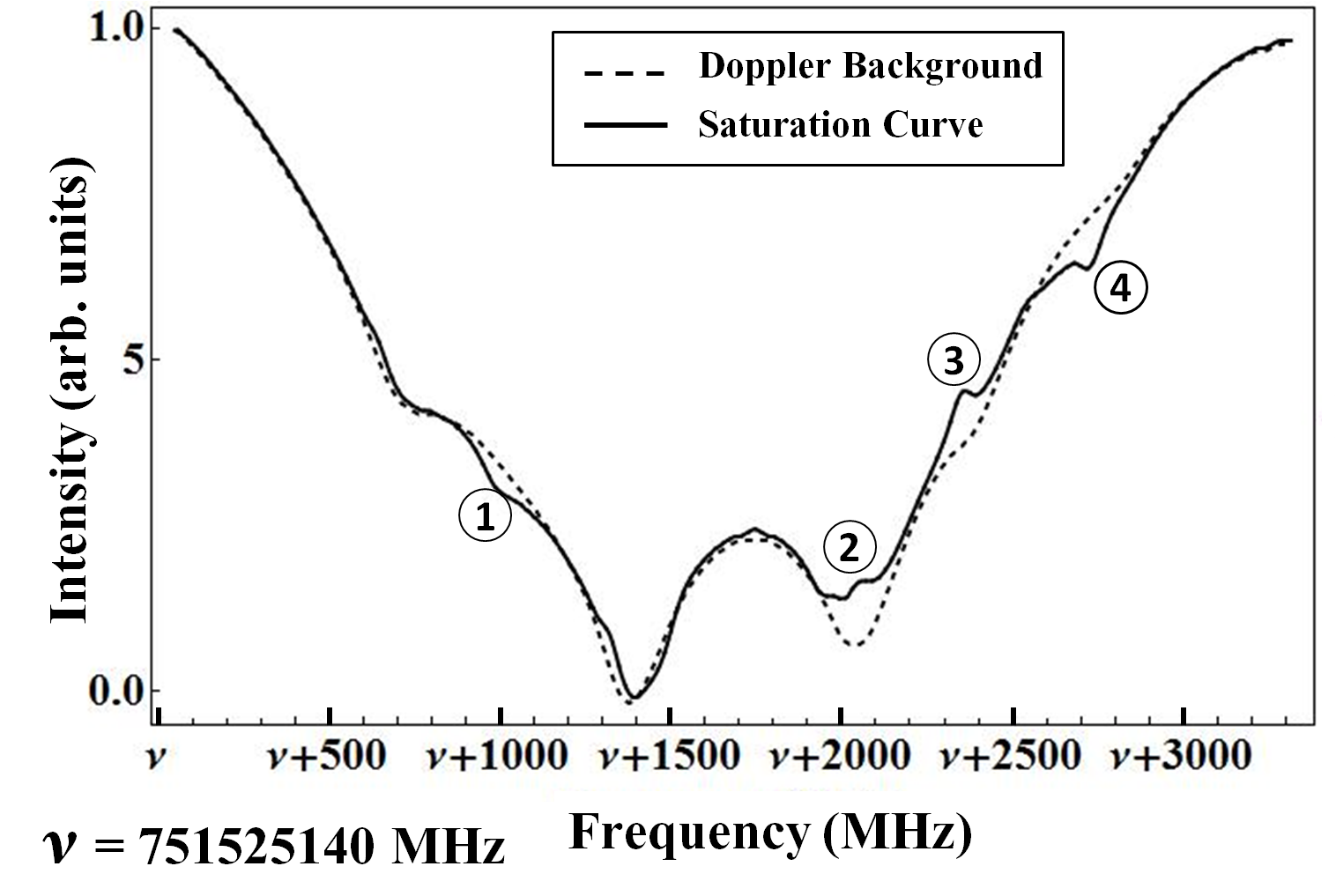}
\caption{$^{1}S_{0}\leftrightarrow{^1}P_{1}$ transition peaks for different isotopes of Yb obtained with saturation absorption spectroscopy. The transitions corresponding to labels 1 to 4 are shown in Table \ref{tab:saturationresults}. The dashed line represents the Doppler background.}
\label{fig:saturation}
\end{figure}

\section{Saturation absorption spectroscopy}
To provide a comparison for the transition frequencies and isotope shifts obtained by our spot method, a typical saturation absorption spectroscopy experiment was set up. The saturation
spectroscopy setup consisted of a pump and probe beam (127 mWcm$^{-2}$ and 2 mWcm$^{-2}$ respectively) counter propagating and overlapping through ytterbium atomic beam. By scanning the wavelength of the external cavity diode laser the saturation plot shown in figure \ref{fig:saturation} was produced. The light gray line represents the Doppler background, which was obtained by scanning the wavelength but blocking the pump beam. With the pump beam included the dark trace was produced. Saturation peaks for different isotopes can be seen at label 1, 2, 3 and 4 on the graph and corresponding frequencies are listed in table: \ref{tab:saturationresults} and compared with the results obtained with the spot method. It can be seen that the results obtained from the saturation spectroscopy are in good agreement with the results from our fluorescence spot method. However the signal to noise ratio using the saturation spectroscopy setup is poor compare to spot method. Further to this, saturation spectroscopy can only be used to determine the resonant transition frequencies but not Doppler shifted transition frequencies in non-perpendicular laser-atomic beam configurations which can be useful in ion trap experiments.

\begin{table*}[!htb]
\caption{The absolute frequencies for $^1{S}_0 \leftrightarrow {^1}P_1$ transition line of Yb isotopes obtained by saturation spectroscopy and the spot technique}
\begin{tabular}{|c|c|c|c|c|}
  \hline
  Peak    & Yb Isotope & Transition                                             & Frequency [THz]           & Frequency [THz]\\
          &              &                                                        & (Saturation)              & (Spot Method)\\ \hline
1         &176          & $^{1}S_{0}\leftrightarrow{^1}P_{1}$                     & 751.52615(60)            &751.52615(60) \\\hline
2         &172$^{*}$    & $^{1}S_{0}\leftrightarrow{^1}P_{1}$                     & 751.52714(120)           &751.52720(120)\\
          &173$^{*}$     & $^{1}S_{0}(F=1/2)\leftrightarrow{^1}P_{1}(F=3/2)$       &    "                     &       "        \\
          &173$^{*}$     & $^{1}S_{0}(F=1/2)\leftrightarrow{^1}P_{1}(F=7/2)$       &    "                     &       "        \\\hline
3         &171          & $^{1}S_{0}(F=1/2)\leftrightarrow{^1}P_{1}(F=3/2)$        & 751.52760(60)             &751.52749(60)\\\hline
4         &170$^{*}$     & $^{1}S_{0}\leftrightarrow{^1}P_{1}$                     & 751.52779(120)            &751.52780(120)\\
          &171$^{*}$     & $^{1}S_{0}(F=1/2)\leftrightarrow{^1}P_{1}(F=1/2)$       &                           &               \\
  \hline
\end{tabular}\\
\emph{* not resolved}
\label{tab:saturationresults}
\end{table*}

\section{Doppler Shifted frequency measurements}

In the previous section the atomic motion was perpendicular to the laser beam propagation when measuring resonant transition frequencies. However, in many experiments the atomic motion is not necessarily perpendicular to the laser beams. In this situation a component of the velocity of the atoms is parallel (or anti parallel) to the laser beam and the frequency of the resonance line becomes Doppler shifted and this Doppler shift depends on the temperature of the atomic oven.
Using our method these Doppler shifted transition frequencies can be measured and average velocity of the atoms in the direction of the reference axis and oven temperature can be determined.\\
\begin{figure}[!b]
\centering
\includegraphics[scale=0.35] {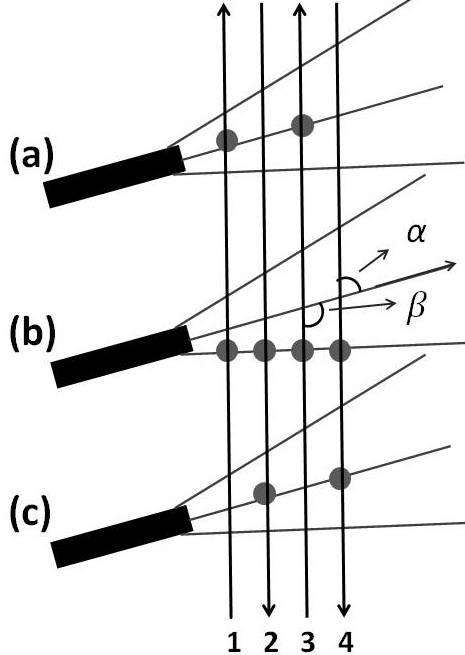}
\caption{Illustration of the counter propagating laser beam pairs making an angle with atomic beam. Laser beam pair (2 and 4) makes an acute angle $\alpha$ and pair (1 and 3) makes  an obtuse angle $\beta$ with the reference axis. To align the pair of spots with the reference axis, laser frequency is blue detuned in (a) and red detuned in (c). In (b) both spot pairs are aligned (but not with the reference axis) when the frequency is at resonance. }
\label{laserbeam}
\end{figure}
\begin{figure}[!b]
  \centering
 \includegraphics[scale=0.118]{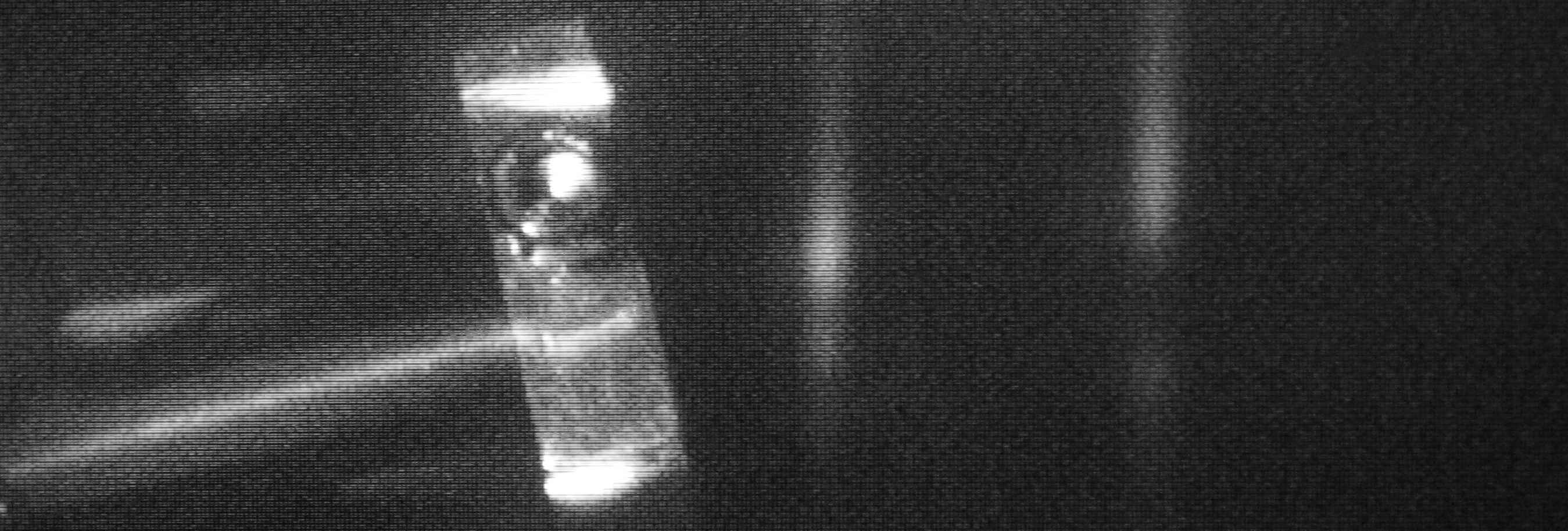} (a)
 \includegraphics[scale=0.12]{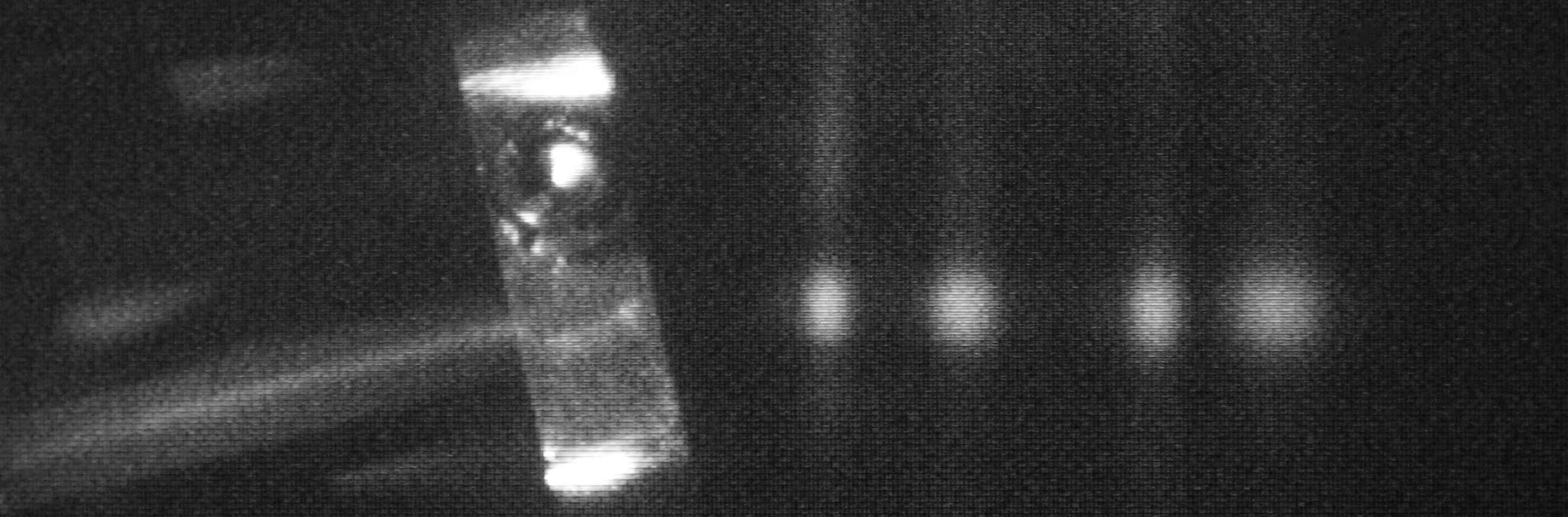} (b)
\includegraphics[scale=0.12]{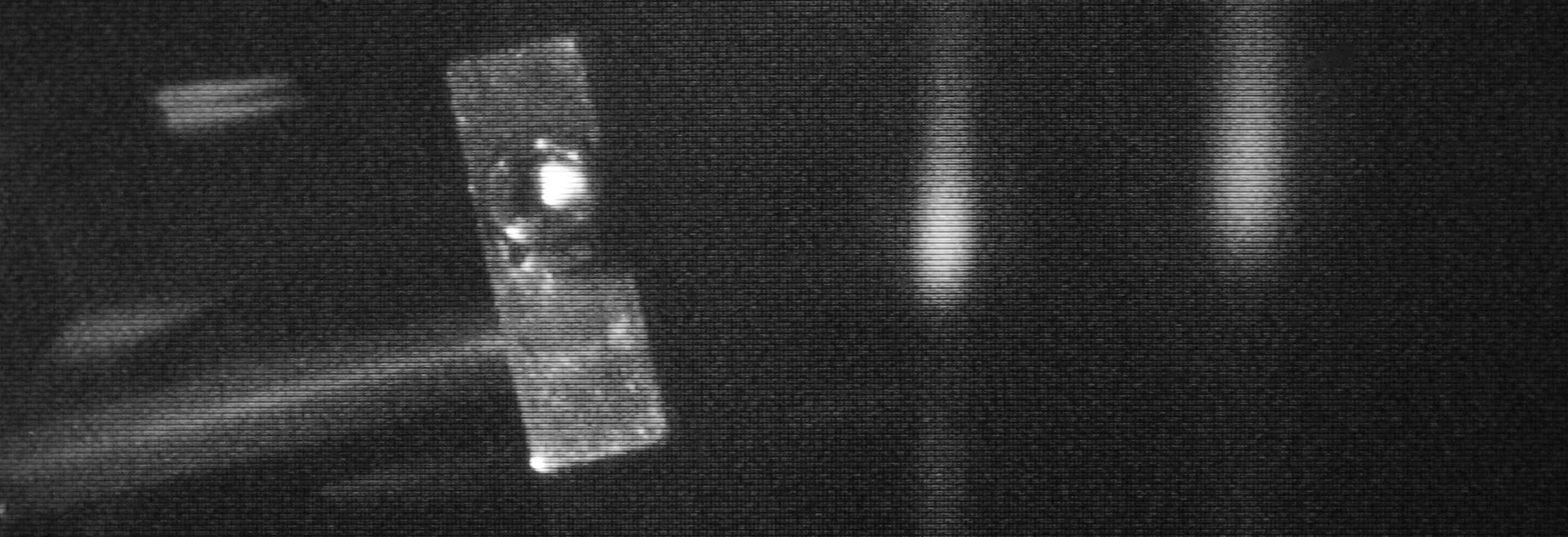} (c)
\caption{Pictures show the atomic oven and fluorescence spots where the laser beams intersect the reference axis at (a) 70$^{o}$ and (c) 110$^{o}$. (b) also demonstrates the non-Doppler shifted resonance where the spots align perpendicular to the laser beams, as seen in the fluorescence spot technique}
\label{angleshift}
\end{figure}
As described earlier when the frequency is detuned from the doppler free resonance two pairs of spots move in opposite directions, relating to the pairs of counter-propagating laser beams. To measure the doppler shifted wavelengths the earlier setup was adjusted by rotating the oven to form an angle with the lasers. This provided maximum atomic flux for the desired angle, while the cylindrical axis of the atomic oven offered a reference axis to which the spots could be aligned. This scheme is illustrated in figure: \ref{laserbeam}. Here the direction of propagation of the laser beams must be taken into account because one pair of the laser beams (2 and 4) makes an acute angle $\alpha$ while the other pair (beam 1 and 3) forms an obtuse angle $\beta$ with the reference axis. In both cases the magnitude of the angular difference from 90$^{o}$ (and hence the frequency shift) is the same, but the sign is different. To align a pair of spots to the reference axis the laser frequency was detuned from resonance by a certain amount. Blue detuning the laser frequency causes fluorescence spots (corresponding to lasers 2 and 4) to align to the reference axis, while red detuning causes fluorescence spots (from lasers 1 and 3) to align to the reference axis.  Figure: \ref{angleshift} demonstrates these Doppler shifted resonance spots when the angle between laser and the atomic beam is (a) $\alpha=70^{o}$ and (c) $\beta=110^{o}$. A particular advantage of the spot method can be realised in figure: \ref{angleshift}(b), that when on resonance the four spots form a line perpendicular to the laser beams. This allows for the resonant transition frequency to be measured independent of the angle between the laser beam and atomic beam. The Doppler shift on the resonance frequency depends on the velocity component, of the atoms, parallel (or anti parallel) to the propagation of the lasers, which is dependant upon the cosine of the angle $\theta$ between the atomic motion and the propagation of laser. Therefore the frequency shift, $\Delta f$, is given by \cite{Book}

\begin{equation}
\label{dopplershift}
\Delta f=\frac{f}{c}v\cos\theta
\end{equation}

\begin{figure}[!t]
\centering
\includegraphics[width=80mm] {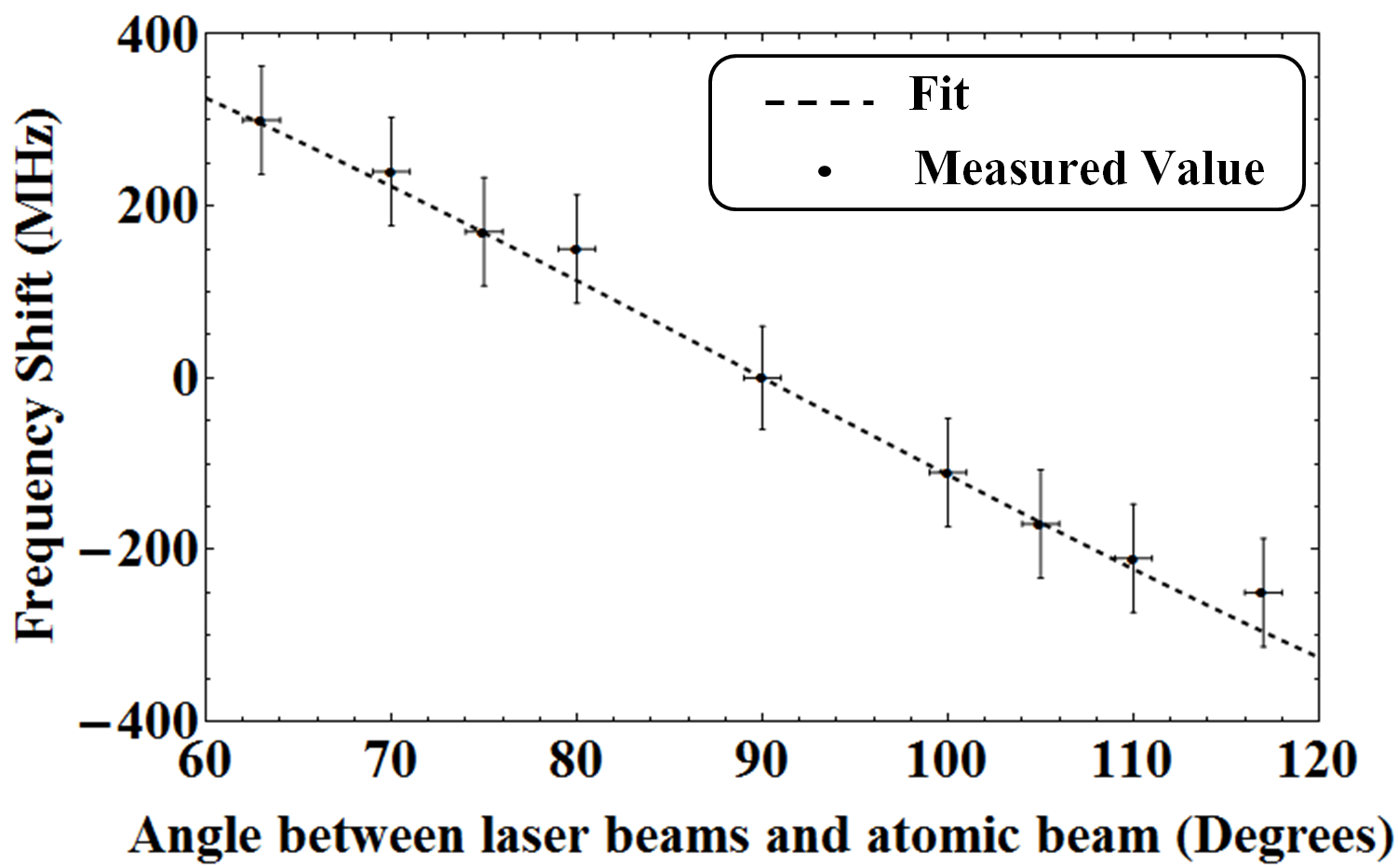}
\caption{Doppler frequency shift in $^{1}S_{0}\leftrightarrow{^1}P_{1}$ transition of $^{174}$Yb as a function of angle between the laser beams and atomic beam}
\label{Angle}
\end{figure}

where ${f}$ is the Doppler free transition frequency, ${c}$ is the speed of light, ${v}$ is the mean velocity of the atoms along the direction of reference axis and $\theta$ is the angle between the atom motion and laser beam. The $v cos\theta$ term represents the mean velocity component parallel to the laser beam.\\
 The reference axis was rotated to make angles of 63, 70, 75 and 80 degrees with one pair of lasers, and hence 117, 110, 105 and 100 degrees with the other pair of lasers. We measured the Doppler shifted frequencies, when the spots aligned with the reference axis, and our measurements of these frequencies are shown in figure \ref{Angle}. Equation \ref{dopplershift} was fitted to the data and enables us to derive the average velocity of the atoms in the direction of atomic beam. Using our data we have determined the average atomic velocity to be 260$\pm$ 20 ms$^{-1}$ and therefore the temperature of the atomic beam can be estimated.\\
Using frequencies measured from our technique we were able to photoionise and trap Yb$^+$ isotopes in an ion trapping experiment \cite{Jim}. To ionise a Yb atom a 399 nm photon is required to drive the $^{1}S_{0}\leftrightarrow{^1}P_{1}$ transition where a further 369 nm photon excites an electron past the continuum \cite{Balzer,braun}. In the ion trapping experiment the atomic oven and 399 nm laser made an angle of 63$\pm$2$^{o}$. We were able to verify the resonance frequencies for the $^{1}S_{0}\leftrightarrow{^1}P_{1}$ transitions by photoionisation of different Yb isotopes.

\section{Conclusion}
We have demonstrated a simple technique to determine the resonant transition frequencies and isotope shifts, of the 398.91 nm $^{1}S_{0}\leftrightarrow{^1}P_{1}$ transition, for different stable Yb isotopes. We present new values for the resonant frequencies that differ from previously published work \cite{Das}. Saturation absorption spectroscopy was also performed and the frequencies measured were in good agreement to those obtained via our florescence spot method. The fluorescence spot method has an advantage over absorption saturation spectroscopy as the former method is easier in setup and provides larger signal to noise ratio. By small modifications to the setup, the fluorescence spot method can also be used to measure the Doppler shifted transition frequencies that occur when the laser and the atomic beam are not perpendicular and allowing for predictions of transition frequencies in realistic ion trap setups. Furthermore, this technique allows us to derive the average thermal velocity of atoms in the direction of atomic beam.\\
\section*{ACKNOWLEDGMENTS}
This work was supported by the UK Engineering and Physical Sciences Research
Council (EP/E011136/1, EP/G007276/1), the European Commission's Sixth
Framework Marie Curie International Reintegration Programme
(MIRG-CT-2007-046432), the Nuffield Foundation and the University of Sussex.

\end{document}